\documentclass[pra,noshowpacs]{revtex4}

\def\var{\varepsilon}

\begin{document}

\title{Stabilizing quantum metastable states\\ in a time-periodic potential}

\author{\underline{Choon-Lin Ho}}
\affiliation{Department of Physics, Tamkang University, Tamsui
25137, Taiwan}
\author{Chung-Chieh Lee}
\affiliation{ Department of Electronic Engineering, Tung Nan
Institute of Technology, Shenkeng, Taipei 222, Taiwan}

\begin{abstract}
In this talk we present a model to demonstrate how time-periodic
potential can be used to manipulate quantum metastability of a
system.  We study metastability of a particle trapped in a well
with a time-periodically oscillating barrier in the Floquet
formalism. It is shown that the oscillating barrier causes the
system to decay faster in general. However, avoided crossings of
metastable states can occur with the less stable states crossing
over to the more stable ones.  If in the static well there exists
a bound state, then it is possible to stabilize a metastable state
by adiabatically increasing the oscillating frequency of the
barrier so that the unstable state eventually cross-over to the
stable bound state. It is also found that increasing the amplitude
of the oscillating field may change a direct crossing of states
into an avoided one. Hence, one can manipulate the stability of
different states in a quantum potential by a combination of
adiabatic changes of the frequency and the amplitude of the
oscillating barrier.
\end{abstract}

\maketitle

\section{Introduction}

Ever since the advent of quantum mechanics, quantum tunneling has
been an important and fascinating subject. This phenomenon arises
frequently in physics.  In fact, one of the first successful
applications of quantum mechanics has been the explanation of the
$\alpha$-decay of atoms as a quantum tunneling process
\cite{alpha}.  Recent examples include tunneling phenomena in
semiconductors and superconductors \cite{SS}, in Josephson
junction systems \cite{JJS}, resonant tunneling in heterojunction
nanostructures \cite{Esaki}, tunneling ionization of atoms
\cite{DK}, photon-assisted tunneling in superconducting junctions
and semiconductor superlattices \cite{PAT}, etc.

In cosmology, quantum metastable states play an essential r\^ole
in some versions of the inflationary models of the early universe
\cite{Guth}. In these models inflation of the early universe is
governed by a Higgs field trapped in a metastable state. Inflation
ends when the metastable state decays to the true ground state of
the universe. During inflation the universe expands exponentially.
It is thus obvious that the metastable state of the Higgs field is
trapped in a rapidly varying potential.  The problem is therefore
a truly time-dependent one. Unfortunately, owing to the inherent
difficulties of the problem, more often than not one has to
consider the decay of the Higgs field in a quasi-stationary
approximation, in which the decay is studied assuming a static
potential \cite{Kolb}. Before one can deal with the decay of Higgs
field in the non-stationary potential, it is desirable to gain
some insights first by studying metastability in time-dependent
potential in simple quantum-mechanical models.

An early attempt at studying the effects of time-varying forces on
quantum metastability appears in Fisher's work \cite{Fisher},
which was motivated by an experiment on quantum tunneling of the
phase in a current-biased Josephson junction with a weak microwave
perturbation \cite{MDC}.  In this work  Fisher considered the
general problem of quantum tunneling in a metastable well with a
weak oscillatory force. There he reformulated the standard WKBJ
approach to quantum decay in order to include a weak
time-dependent perturbation.  For a class of metastable potentials
which interpolates between the cubic potential and a truncated
harmonic-oscillator potential, he showed that the decay rate is
generally enhanced by the weak oscillatory force.  The potential
considered by Fisher has a number of oscillator-like levels near
its minimum.  The opposite situation where only two levels are
present was considered by Sokolovski \cite{Sokolovski}, who
studied the effect of a small AC field mixing two levels in the
well on the tunneling rate in a semiclassical framework.

The results in \cite{Fisher,Sokolovski} are quite general for a
class of weak oscillatory forces.  However, it is desirable to
consider other possibilities, e.g., exact solutions and/or
nonperturbative results.

In \cite{HoLee,LeeHo} we have considered  simple driven quantum
metastable models in which a particle is trapped in a well with a
periodically driven rectangular barrier. In order to do away with
any restriction of amplitude or frequency of the driving force,
and of the number of states in the potential, we treat the
problems in the framework of the Floquet formalism
\cite{Shirley,Sambe,LR}.  An exact expression determining the
Floquet quasi-energies of stable/metastable states in the well is
derived. From the solution of this equation we find that while the
oscillating barrier makes the system decay faster in general,
there is the possibility that avoided crossings of metastable
states can occur with the less stable states crossing over to the
more stable ones.

That an oscillating potential can affect the tunneling property of
a system has also been noticed before, eg. in quantum transport
process \cite{LR,CDT}). In \cite{LR} it was demonstrated that an
propagating particle at appropriate incident energy can be trapped
into a bound state by an oscillating square-well.  In \cite{CDT}
it was found that a particle can be localized in one side of a
time-dependent double well if the amplitude and the frequency of
the oscillating field were chosen properly.  Our examples shows
how a time-periodic field can modify the metastability of a
decaying state.

\section{The model}

To be concrete, in this talk we present the model considered in
\cite{HoLee}.  This model consists of a particle of mass $m$
trapped in a square well with a harmonically oscillating barrier,
\begin{eqnarray}
 V(x,t)=\left\{
  \begin{array}{llll}
    \infty~, & x<0~, \\
    0~, & 0\leq x<a~,\\
    V_{0}+V_{1}\cos(\omega t)~, & a\leq x\leq b~, \\
    V_0^\prime ~, & x>b~.
  \end{array} \right.
  \label{potential}
\end{eqnarray}
Here $V_0,V_1, V_0^\prime$ and $\omega$ are positive parameters,
with $V_0^\prime <V_0$ and $V_1<V_0-V_0^\prime$. According to the
Floquet theorem, the wave function of a time-periodic system has
the form $\Psi_{\var}(x,t)=e^{-i\var t/\hbar}\Phi_{\var}(x,t)$,
where $\Phi_{\var}(x,t)$ is a periodic function with the period
$T=2\pi/\omega$, i.e. $\Phi_{\var}(x,t+T)=\Phi_{\var}(x,t)$, and
$\var$ is the Floquet quasi-energy, which we will call Floquet
energy for brevity. It should be noted that the Floquet energy is
determined only modulo $\hbar\omega$. For if $\{\var, \Phi_\var\}$
is a solution of the Schr\"odinger equation, then
$\{\var^\prime=\var+n\hbar\omega~, \Phi_{\var^\prime}=\Phi_\var
\exp(in\omega t)\}$ is also a solution for any integer $n$.  But
they are physically equivalent as the total wave function
$\Psi_\var$ is the same \cite{Sambe}. All physically inequivalent
states can be characterized by their reduced Floquet energies in a
zone with a width $\hbar\omega$. We therefore consider solutions
of $\var$ only in the first Floquet zone, i.e. $\var\in
[0,\hbar\omega)$.

The wave function of the particle is described as follows
\cite{HoLee}:
\begin{eqnarray}
 \Psi(x,t)&=& e^{-i\var t/\hbar}\Phi_\var (x,t)\\ \nonumber
 &=& e^{-i\var t/\hbar}\left\{
  \begin{array}{lll}
     \sum_{n=-\infty}^\infty A_{n}\sin(k_{n}x)
    e^{-in\omega t}~, & 0\leq x<a~, \\
   \sum_{n=-\infty}^\infty\sum_{l=-\infty}^\infty
    \left(a_{l}e^{q_{l}x}+b_{l}e^{-q_{l}x}\right)
 J_{n-l}\left(\frac{V_1}{\hbar\omega}\right)e^{-in\omega t}~, & a\leq x\leq b~,
\\    \sum_{n=-\infty}^\infty t_{n}e^{ik'_{n}x}e^{-in\omega t}~, &
x>b~,
  \end{array} \right.
  \label{wf}
\end{eqnarray}
where
\begin{eqnarray}
 k_{n}&=&\sqrt{2m(\var+n\hbar\omega)}/\hbar~, \nonumber\\
q_{l}&=&\sqrt{2m(V_0-\var-l\hbar\omega)}/\hbar~,\label{k}\\
k'_{n}&=&\sqrt{2m(\var+n\hbar\omega-V'_{0})}/\hbar~, \nonumber
\end{eqnarray}
 and $J_n$'s are
the Bessel functions.  In the region $x>b$, we have adopted
Gamow's outgoing boundary condition, namely, there is no particle
approaching the barrier from the right \cite{alpha}. Matching the
wave function and its first derivative at the boundaries $x=a$ and
$x=b$, we obtain the relations among the coefficients
$A_{n},a_{n},b_{n}$ and $t_{n}$:
\begin{eqnarray}
 A_{n}\sin(k_{n}a) &=&
 \sum_{l}\left(a_{l}e^{q_{l}a}+b_{l}e^{-q_{l}a}\right)J_{n-l}
 (\alpha)~,
 \nonumber
 \\
 k_{n}A_{n}\cos(k_{n}a) &=&
 \sum_{l}q_{l}\left(a_{l}e^{q_{l}a}-b_{l}e^{-q_{l}a}\right)
 J_{n-l}(\alpha)~,
 \label{b.c.1}
 \\
 t_{n}e^{ik'_{n}b} &=&
 \sum_{l}\left(a_{l}e^{q_{l}b}+b_{l}e^{-q_{l}b}\right)J_{n-l}
 (\alpha)~,
 \nonumber
 \\
 ik'_{n}t_{n}e^{ik'_{n}b} &=&
 \sum_{l}q_{l}\left(a_{l}e^{q_{l}b}-b_{l}e^{-q_{l}b}\right)
 J_{n-l}(\alpha)~,
 \nonumber
\end{eqnarray}
where $\alpha\equiv V_{1}/\hbar\omega$. The Floquet energy is
determined from these relations by demanding non-trivial solutions
of the coefficients.  In practice, however, we must truncate the
above equations to a finite number of terms, or sidebands as are
usually called in the literature, eg. $n=0, \pm1,\ldots,\pm N$ .
The number $N$ is determined by the frequency and the strength of
the oscillation as $N>V_{1}/\hbar\omega$ \cite{LR}.

By demanding non-trivial solutions of the coefficients $a_{n}$,
$b_{n}$, $A_n$ and $t_{n}$ in  Eq.(\ref{b.c.1}), we obtain, after
some tedious algebra, an equation which determines the Floquet
energy $\var$:
\begin{eqnarray}
 F_{4}\frac{q_{0}}{k_{0}}\tan k_{0}a +F_{2}
 =\frac{F_{8}q_{0}+iF_{6}k'_{0}}{F_{7}q_{0}-iF_{5}k'_{0}}
 \left(F_{3}\frac{q_{0}}{k_{0}}\tan k_{0}a -F_{1}\right)
 e^{-2q_{0}(b-a)}~.
 \label{solution}
\end{eqnarray}
Here $F_i (k_0, k_0^\prime,\omega, V_1)$ are functions of $k_0,
q_0, k'_0$ (see \cite{HoLee} for their complete forms), and hence
are connected with the Floquet energy $\var$ (c.f. Eq.~(\ref{k})).
If the solutions $\var$ of Eq.(\ref{solution}) are complex (real)
numbers, the corresponding Floquet states are metastable (stable)
states. The non-decay probability $P(t)$, which is the probability
of the particle still being trapped by the potential barrier at
time $t>0$, is given by
\begin{eqnarray}
 P(t) &=& \frac{\int_{0}^{b}\left|\Psi(x,t)\right|^2dx}
 {\int_{0}^{b}\left|\Psi(x,0)\right|^2dx}
 \nonumber \\
 &=& e^{2Im(\var)t/\hbar} \frac{\int_{0}^{b}\left|\Phi_\var(x,t)\right|^2dx}
 {\int_{0}^{b}\left|\Phi_\var(x,0)\right|^2dx}
 \label{P(t)}\\
 &\equiv & e^{2Im(\var)t/\hbar} h(t)~,\nonumber
\end{eqnarray}
with $P(0)=1$. The imaginary part of the Floquet energy, which
enters $P(t)$ via the factor $\exp(2Im(\var)t/\hbar)$, gives a
measure of the stability of the system.  Unlike the static case,
however, here $P(t)$ is not a monotonic function of time, owing to
the time-dependent function $h(t)$ after the exponential factor in
Eq.({\ref{P(t)}). But since $h(t)$ is only a periodic function
oscillating between two values which are of order one, the
essential behavior of $P(t)$ at large times is still mainly
governed by the exponential factor.  Hence, as a useful measure of
the non-decay rate of the particle in the well, we propose a
coarse-grained non-decay probability $\bar{P}(t)$ defined as
\begin{eqnarray}
\bar{P}(t)\equiv  e^{2Im(\var)t/\hbar} \langle h(t) \rangle~,
\label{bP(t)}
\end{eqnarray}
where $\langle h(t) \rangle$ is the time-average of $h(t)$ over
one period of oscillation (some graphs of $\bar{P} (t)$ are
presented in \cite{HoLee}).

The coefficients $F_{i}(k_{0},k_0^\prime,\omega,V_{1})$ all
approach to one in the limit $\alpha=V_1/\hbar\omega\to 0$.  Hence
in the limit $V_{1}\to 0$ or $\omega\to \infty$,
Eq.(\ref{solution}) reduces to the corresponding equation for the
case of static potential with potential $V_0$ in the region $a\leq
x \leq b$, and the Floquet energy in this limit is just the (real
or complex) eigenenergy of the static case. This is
understandable, since in the limit $V_1\to 0$ the potential
becomes static, and at high frequencies the particle in the well
will only see a time-averaged barrier of effective height $V_0$.

\section{Numerical results}

We now study numerical solutions of Eq.(\ref{solution}) with a
specific potential.  We take $a=1,~b=2,~V_0=15$ and
$V_0^\prime=V_0/2$ in the atomic units (a.u.) ($e=m_e=\hbar=1$).
In the static case this potential supports one bound state, with
energy $E_0/V_0=0.232123$,  and one metastable state, with complex
energy $E_1/V_0=0.864945-0.00255261i$.  For the oscillating
potential, we solve Eq.(\ref{solution}) in 2-sideband
approximation, i.e. we take $N=2$.  This is accurate enough for
oscillating frequency $\omega\geq V_1/2$.

Fig.~1 and 2 present the graphs of the real and imaginary parts of
the Floquet energy ($\var/V_0$) as a function of $\omega/V_0\geq
0.2$ with $V_1=0.1V_0$ and $0.2V_0$, respectively. We find that
the solutions of Eq.(\ref{solution}) have the form $\var=\var_0 +
n\omega$ ($n=0,\pm 1, \pm 2,\ldots$), with $\Re(\var_0)$ (the
horizontal branch) lie close to the energies $E_0$ and $\Re(E_1)$
in the static potential. That is, these branches of $\Re(\var)$
emanate from either $E_0$ or $\Re (E_1)$ at $\omega=0$. Branches
emerging from the same point have the same imaginary part.
Numerical results show that, with the barrier oscillating, the
stable state ($E_0$) in the static case becomes unstable, and the
unstable state ($E_1$) will decay even faster. For simplicity, in
Fig.~(1a) and (2a) we show only six branches ($n=0,\pm 1, \pm 2$
and $-3$) emerging from $\Re (E_1)$, and only the central branch
($n=0$) and a subband ($n=-1$) from $E_0$.  As mentioned before,
we only take solutions in the first Floquet zone, $\Re(\var)$
(modulo $\omega$), which are points under the line
$\Re(\var)=\omega$.

From Fig.~1 and 2 we also see that a direct crossing occurs at
frequency $\omega\approx \Re(E_1 - E_0)/2$ (point $c$). However,
independent of the values of $V_1$, as $\omega$ approaches the
frequency $\omega\approx \Re(E_1 - E_0)=0.632822V_0$, an avoided
crossing ($e,e^\prime$) between the real parts of the Floquet
energies occurs. Fig.~2 indicates that larger values of $V_1$ only
enhance the instability of the system and the repulsion between
the two levels at avoided crossing. Thus as the frequency $\omega$
is increased, the state emanating from $E_1$ has Floquet energy
with real part given by values along the path $abb^\prime
cdd^\prime e^\prime f$ (the dark dotted curve), while the real
part of Floquet energy of the state emerging from $E_0$ lies along
the path $cegg^\prime h$ (the solid curve).  The imaginary parts
of these two paths are depicted in Fig.~(1b) and (2b).  One sees
that an exchange of the imaginary parts takes place at the avoided
crossing $ee^\prime$.  Beyond the avoided crossing, the upper
state becomes more stable than the lower state.  This gives the
possibility of stabilizing an unstable state by an oscillating
field.  We recall that as $\omega\to \infty$, Eq.(\ref{solution})
reduces to the one for the static potential. In the example
considered here, the lower state supported by the static well is a
stable bound state, and hence the unstable upper Floquet state can
be made stable in the high frequency limit.  Even more simply, the
same aim can be achieved by adiabatically tuning down the
amplitude $V_1$ just after the avoided crossing, as in this limit
the potential becomes the static one.

Finally, it is interesting to note here the r\^ole of amplitude
$V_1$ of the oscillating barrier in the model (a detailed study is
presented in \cite{LeeHo} . As we have seen, the presence of a
non-vanishing $V_1$ always makes the system less stable.  However,
if $V_1$ is reduced, an avoided crossing may turn into a direct
one. In the present case, the avoided crossing $ee'$ changes into
a direct crossing for $V_1/V_0<0.03$. Conversely, increasing $V_1$
could change a direct crossing into an avoided crossing.  At an
avoided crossing, the imaginary parts of the Floquet energies
cross, while the real parts do not. At a direct crossing, it is
the real parts, not the imaginary parts, that cross. But the more
(less) stable state has the tendency to become less (more) stable.
This is evident from the Floquet energy at the direct crossing
point $c$ in Fig.~1 and 2.  These observations are consistent with
the semiclassical results obtained in \cite{Sokolovski} by
perturbative methods. Hence, by a combination of adiabatic changes
of the frequency and the amplitude of the oscillating barrier, one
can manipulate the stability of different states in a quantum
potential: tune up $V_1$ until a direct crossing becomes an
avoided one, increase $\omega$ so that the avoided crossing is
passed, then reduce $V_1$ to make the potential static. In the
process, two states in the well are interchanged.

\section{summary}

To summarize, our results show that an oscillating potential
barrier generally makes a metastable system decay faster. However,
the existence of avoided crossings of metastable states can switch
a less stable state to a more stable one.  If in the static well
there exists a bound state, then it is possible to stabilize a
metastable state by adiabatically changing the oscillating
frequency and amplitude of the barrier so that the unstable state
eventually cross-over to the stable bound state.  Thus an
time-dependent potential can be used to control the stability of a
particle trapped in a well.

The model we present here has its barrier oscillating.  One may
also consider the same model with an oscillating bottom instead.
However, this latter model turns out to have identical Floquet
energy spectrum as the one presented here for the same frequency
and amplitude of the oscillating field.  This can be explained by
a discrete transform which connects the equations of boundary
conditions of the two models, and by a gauge transformation which
maps the wave functions of the two models \cite{LeeHo}.

\vskip 2cm

\begin{acknowledgments}
 This work was supported in part by the National Science Council of
the Republic of China through Grant No. NSC 94-2112-M-032-007.
\end{acknowledgments}

\vskip 3truecm
\centerline{\bf Figure Captions}
\begin{description}
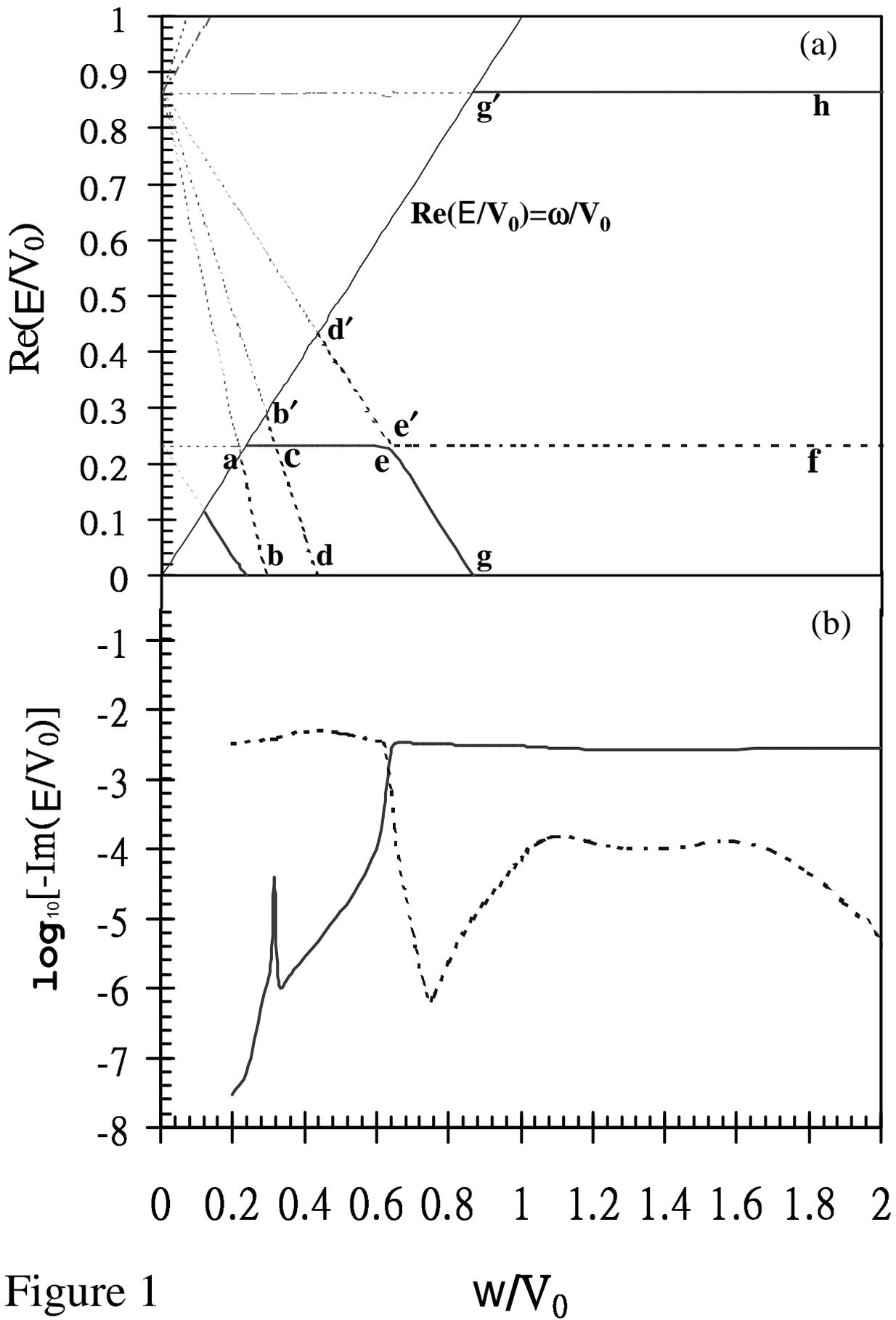
\item[Figure 1.]  The Floquet
energies ($\var/V_0$) of the two metastable states versus the
barrier oscillating frequency ($\omega/V_0$) for $V_0=15 a.u.,
V_0^\prime=V_0/2$ and $V_1/V_0=0.1$ in the atomic units (a.u.)
($e=m_e=\hbar=1$) . In (a) the real parts of the Floquet energies
are shown in the first Floquet zone under the line
$\Re(\var)=\omega$ (the straight line). The light dotted lines
show how the different branches emanate from the two states in the
static case (with $E_0/V_0=0.232123$, and
$\Re(E_1)/V_0=0.864945$).  In (b) the corresponding imaginary
parts of the Floquet energies of the two states are plotted. The
dotted curve corresponds to the state with real parts given along
the path $abb^\prime cdd^\prime e^\prime f$, and the solid curve
corresponds to the state with real parts given along $cegg^\prime
h$.

\vskip 1cm

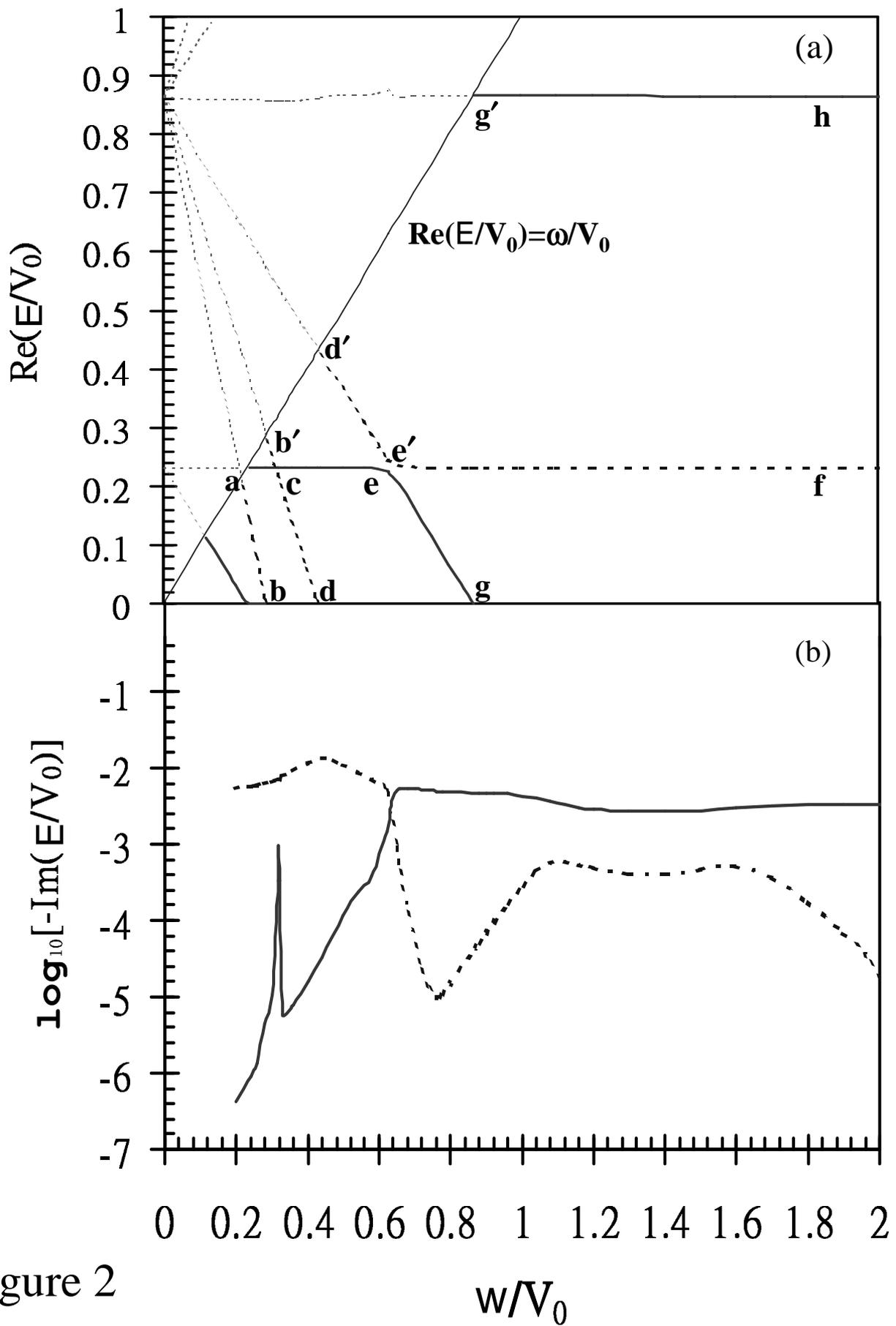
\item[Figure 2.]  Same plot as Fig.~1 for $V_0=15 a.u.,
V_0^\prime=V_0/2$ and $V_1/V_0=0.2$.

\end{description}


\newpage

\begin{thebibliography}{9}
\bibitem{alpha} G.Gamow, \emph{Z. Phys.} \textbf{51}, 204 (1928);
R.W. Curney and E.U. Condon, \emph{Phys. Rev.} \textbf{33}, 127
(1929).
\bibitem{SS} For overviews on these developments, see eg.:  L. Esaki, \emph{Proc.
IEEE} \textbf{62}, 825 (1974); I. Giaever, \emph{Science}
\textbf{183}, 1253 (1974); B.D. Josephson, \emph{ibid.}
\textbf{184}, 527 (1974).
\bibitem{JJS} S. Washburn, R.A. Webb, R.F. Voss, and S.M. Faris, \emph{Phys.
Rev. Lett.} \textbf{54}, 2712 (1985); D.B. Schwarz, B. Sen, C.N.
Archie, and J.E. Lukens, \emph{ibid.} \textbf{55}, 1547 (1985);
M.H. Devoret, J.M. Martinis, and J. Clarke, \emph{ibid.}
\textbf{55}, 1908 (1985).
\bibitem{Esaki} R. Tsu and L. Esaki, \emph{Appl. Phys.  Lett.}
\textbf{22}, 562 (1973); L.L. Chang, L. Esaki and R. Tsu,
\emph{ibid.} \textbf{24}, 593 (1974).
\bibitem{DK} For a recent review, see eg.: N.B. Delone and V.P. Krainov,
\emph{Multiphoton Processes in Atoms}, Springer-Verlag, Berlin
Heidelberg, 1994.
\bibitem{PAT} A.H. Dayem and R.J. Martin, \emph{Phys. Rev. Lett.} \textbf{8}, 246
(1962); P.K. Tien and J.P. Gordon, \emph{Phys. Rev.} \textbf{
129}, 647 (1963); B.J. Keay et al., \emph{Phys. Rev. Lett.}
\textbf{ 5}, 4098 (1995).
\bibitem{Guth} A.H. Guth, \emph{Phys. Rev.} \textbf{D23}, 347 (1981);
D. La and P.J. Steinhardt, \emph{Phys. Rev. Lett.} \textbf{62},
376 (1989).
\bibitem{Kolb} E.W. Kolb and M.S. Turner, \emph{The
Early Universe}, Addison-Wesley, New York, 1990.
\bibitem{Fisher} M.P.A. Fisher, \emph{Phys. Rev.} \textbf{ B37}, 75 (1988).
\bibitem{MDC} J.M. Martinis, M.H. Devoret, and J. Clarke, \emph{Phys. Rev.
Lett.} \textbf{ 55}, 1543 (1985).
\bibitem{Sokolovski} D. Sokolovski, \emph{}Phys. Lett.  \textbf{ A132}, 381
(1988).
\bibitem{HoLee} C.L. Ho and C.C. Lee, \emph{Phys. Rev.} \textbf{ A71},
012102 (2005).
\bibitem{LeeHo} C.C. Lee and C.L. Ho, \emph{Ann. Phys.} \textbf{320},
175 (2005).
\bibitem{Shirley} J.H. Shirley, \emph{Phys. Rev.} \textbf{ 138}, B979 (1965).
\bibitem{Sambe} H. Sambe, \emph{Phys. Rev.} \textbf{ A7}, 2203 (1973).
\bibitem{LR} W. Li and L.E. Reichl, \emph{Phys. Rev.}\textbf{B60},
15 732 (1999).
\bibitem{CDT} F. Grossmann, T. Dittrich, P. Jung, and P. H\"anggi, \emph{Phys. Rev.
Lett.} \textbf{ 67}, 516 (1991).

\end{thebibliography}
\end{document}